\documentclass[aps,pre,floatfix,superscriptaddress,twocolumn]{revtex4-1}

\usepackage{graphicx,psfrag,color,amsmath,bbm}
\usepackage{bm}
\newcommand{\average}[1]{\langle{#1}\rangle}

\newcommand{\E}{\mathrm{e}}
\newcommand{\D}{\mathrm{d}}

\newcommand{\ii}{\mathbbm{i}}

\newcommand{\vphi}{\varphi}

\newcommand{\bFu}{\bar F_1}
\newcommand{\bFd}{\bar F_2}

\newcommand{\p}[1]{\left({#1}\right)}
\newcommand{\pq}[1]{\big[{#1}\big]}
\newcommand{\pg}[1]{\big\{{#1}\big\}}

\newcommand{\alb}[1]{#1}

\newcommand{\Xo}{x^{(0)}}
\newcommand{\Yo}{y^{(0)}}

\newcommand{\Xu}{x^{(1)}}
\newcommand{\Yu}{y^{(1)}}

\newcommand{\qo}{q^{(0)}}

\newcommand{\hqo}{\bar q^{(0)}}
\newcommand{\Qo}{Q^{(0)}}
\newcommand{\qu}{q^{(1)}}
\newcommand{\qd}{q^{(2)}}

\usepackage{comment}
\includecomment{mycomments}

\begin{document}

\title{Quantum duets working as autonomous thermal motors}

\author{Michael Drewsen}
\author{Alberto Imparato}

\affiliation{Department of Physics and Astronomy, University of Aarhus\\ Ny Munkegade, Building 1520, DK--8000 Aarhus C, Denmark}

\date{}

\begin{abstract}
We study the dynamic properties of a thermal autonomous machine made up of two quantum Brownian particles, each of which is in contact with an environment at different temperature and moves on a periodic sinusoidal track. When such tracks are shifted, the center of mass of the system exhibits a non-vanishing velocity, for which we provide an exact expression in the limit of small track undulations.
We discuss the role of the broken spatial symmetry in the emergence of directed motion in thermal machines. We then consider the case in which  external deterministic forces are applied to the system, and characterize its steady state velocity. 
If the applied  external force opposes  the system motion, work can be extracted from such a steady state thermal machine, without any external cyclic protocol.
When the two particles are not interacting, our results reduce to those of refs.~\cite{Fisher85,Aslangul87} for a single particle moving in a periodic tilted potential.
We finally use our results for the motor velocity to check the validity of the quantum molecular dynamics algorithm  in the non--linear, non--equilibrium regime.
\end{abstract}
\maketitle

Emerging technologies  have been used  to engineer and realise quantum devices  which are the nanoscopic counterparts of heat engines  \cite{Rossnagel16} or thermoelectric
transducers \cite{Thierschmann2015}.
Similarly to their macroscopic analogues, 
quantum engines are open systems that exchange energy with the  surrounding environments in form of work and heat, and the study of their dynamic and thermodynamic properties  has attracted considerable interest \cite{Devoret2014,Kosloff13}
in an effort to extend the concepts of the classical thermodynamics to the quantum regime \cite{Anders16,Alicki18}.

The properties of quantum reciprocating motors have been extensively investigated. Such motors are characterized by a time dependent Hamiltonian, and interact cyclically  with baths at different  temperatures, so as to perform Carnot,
Otto or Stirling cycles \cite{Kosloff13, Gelbwaser13, Gelbwaser13a, Insinga16,Alicki18} from which work can be extracted.

Autonomous motors are based on a fundamentally different design: these machines can operate in steady state conditions without any external agent changing their Hamiltonian, or their heath bath.
Within this framework, 
a few works have recently appeared where the dynamic and thermodynamic properties of quantum rotors have been studied and  where the sole  driving force is a thermal gradient \cite{Mari15,Roulet17,Seah18,Fogedby18,Roulet2018,Hovhannisyan_2019}. In such models the heat current between different baths is converted into rotational motion.

The properties of autonomous motors exhibiting directed particle transport on rectilinear tracks have not received much attention in the past, with the noticeable exception of refs.~\cite{Fisher85,Aslangul87}, whose authors studied the steady state velocity of a quantum Brownian particle in a periodic potential under the effect of a constant external force that explicitly breaks the spatial symmetry. More recently ref. \cite{Bissbort17}
studied the conversion of energy current flowing between different baths into motion of a particle on a ring.

In this paper we put forward a rigorous approach to investigate the dynamic properties of a quantum  autonomous thermal motor fuelled by a temperature gradient, and by spatial broken symmetry.
Specifically, we propose a minimal model of quantum motor, which is based on the directed transport emerging in systems with both broken spatial symmetry and thermal equilibrium.  
This model  motor, is the quantum counterpart of the classical  system considered in \cite{Fogedby17}, and consists of two particles sitting in two periodic shifted potentials,  interacting through a third potential, and kept at different temperatures.

We use the Caldeira-Legget model \cite{Caldeira83a,Caldeira83b,Ford88} for the two  heat baths, which allow us to write the dynamic equations for the particles' coordinates  in the Heisenberg representation.
In the limit of small undulations in the periodic potentials, the problem can be solved analytically, and the system steady state velocity can be obtained by studying the motion of a free quantum Brownian particle, and of a Brownian quantum oscillator.
This choice for the baths does not require any assumption on the strength of the system-bath interaction, and allows us to derive our results for baths with an arbitrary distribution  of internal modes.
This is relevant in the context of systems interacting with non-Markovian baths, that has attracted considerable  attention in the physics  community working on dynamics of open quantum systems \cite{Vega17,Nazir18}.
Indeed, while it is commonly believed that non-Markovian behaviour emerges in systems which interact strongly with baths with a structured spectral density \cite{Vega17}, it has been recently shown that non-Markovian behaviour can emerge in systems as simple as a single harmonic oscillator in contact with a Ohmic bath, provided that the bath and the system are initially correlated \cite{Strasberg18}.
\alb{Furthermore, the study of quantum heat machines in the strong coupling regime  has received considerable attention in recent years \cite{Gelbwaser15,Uzdin16,Newman17}.}

We  discuss the case in which deterministic external forces are applied to the autonomous motor. 
The case of constant external force is particularly interesting, as a force counteracting the system center-of-mass motion can extract work from the motor while it operates in steady state conditions without any cyclic manipulation protocol, at variance with the setups characterizing the reciprocating engines, as described, e.g., in \cite{Alicki18}.

We finally use the results obtained for the motor steady state velocity to check the validity of the quantum molecular dynamics (QMD) algorithm in the non-linear, out-of-equilibrium, multi-bath regime.


\section{The model}
\label{model:sec}
In this section we introduce the autonomous motor model as the quantum conterpart of the classical model introduced in \cite{Fogedby17}.
The system Hamiltonian reads 
\begin{eqnarray}
H_0(\{P_i\},\{Q_i\})&=&\frac {P^2_1}{2 m}+  \frac{P^2_2}{2 m}+V(Q_1,Q_2),\label{H0:eq}\\
V(Q_1,Q_2)&=&U(Q_1-Q_2) + V_1(Q_1) +V_2(Q_2),
\label{V:eq}
\end{eqnarray} 
with $V_1$ and $V_2$ two periodic potentials and $U$ an interaction potential.
In the following we will take
\begin{eqnarray}
V_1(Q_1)&=&-V_0 \cos(b Q_1),\label{V:cos} \\
V_2(Q_2)&=&-V_0 \cos(b Q_2+\vphi)\label{V:sin}.
\end{eqnarray} 
As discussed in \cite{Fogedby17}, when one takes the interaction potential to be $U(x_1-x_2)=-k \cos[b (x_1-x_2)]$,  eq.~(\ref{H0:eq}) becomes the Hamiltonian for the $xy$-model in an external field, describing the elastic free energy in ferromagnetic or liquid-crystal systems \cite{Chaikin}.
Throughout this paper we will use the expression 
\begin{equation}
U(Q_1-Q_2)=\frac{k}{2}(Q_1-Q_2)^2,  \label{U:quad}
\end{equation} 
 as it will allow us to obtain closed results for the dynamics when $V_0=0$, as detailed below.
The Hamiltonian  eq.~(\ref{H0:eq}) breaks the spatial symmetry in the sense that when $\varphi\neq 0,\pi$ there is no unilateral translation $\Delta$  such that $V(-Q_1,-Q_2)=V( Q_1+ \Delta,Q_2)$ $\forall Q_1,Q_2$, (or similarly for $Q_2$,  $V( -Q_1,-Q_2)=V( Q_1,Q_2+ \Delta)$).
This broken symmetry is the key ingredient for the system to exhibit directed transport, as discussed below.

We model the thermal baths as ensembles of coupled harmonic oscillators \cite{Caldeira83a,Caldeira83b,Ford88}, and thus the total Hamiltonian (system+bahts)  reads
\begin{equation}
H=H_0+ H_1+H_2
\end{equation} 
with
\begin{equation}
H_i=\sum_k\p{\frac{\mathcal{P}_{k,i}^{2}}{2m_{k,i} }+
\frac{1}{2}m_{k,i}\omega_{k,i}^2(\mathcal{X}_{k,i}-Q_i)^2}
\end{equation} 
 and  $i=1,\,2$, and where $\mathcal{X}_{k,i}$ and $\mathcal{P}_{k,i}$ are the bath variables.
While every harmonic oscillator in each bath is characterized by the frequency $
\omega_{k,i}$ and the mass $m_{k,i}$, the details of the coupling between the baths and the system are embedded in these two quantities \cite{Ford88}.
By introducing the density of states 
\begin{eqnarray}
N_i(\omega)=2\pi\sum_k m_{k,i}\omega_{k,i}^2\delta(\omega-\omega_{k,i}),
\label{aDOS}
\end{eqnarray}
we can treat the spectrum for each bath as continuous. 
In this case, the memory function $\eta_i(t)$ is defined as %
\begin{eqnarray}
&&\eta_i(t)=\theta(t)\int\frac{d\omega}{2\pi} N_i(\omega)\cos\omega t.
\label{aeta1}
\end{eqnarray}

Following the procedure described in \cite{Caldeira83a,Caldeira83b}, one obtains the quantum Langevin equations 
\begin{eqnarray}
m\ddot Q_i=-\partial_i V -\int^t_{t_0} \eta_i(t-t')\dot Q_i(t') \D t'+\xi_i(t) ,
\label{aqlan}
\end{eqnarray}
which are the dynamical equations for the operators $Q_i$ in the Heisenberg picture. The interaction with the baths are now embodied by the quantum noise operators $\xi_i(t)$.
For the specific choice of the potential (\ref{V:eq})--(\ref{V:sin}), one obtains
\begin{eqnarray}
m\ddot Q_1&=&-\int^t_{t_0} \eta_1(t-t')\dot Q_1(t') \D t'\nonumber \\
&& -V_0 b \sin b Q_1-k (Q_1-Q_2)+\xi_1,\label{lang1a}\\
m\ddot Q_2&=&-\int^t_{t_0} \eta_2(t-t')\dot Q_2(t') \D t'\nonumber \\
&& -V_0 b \sin (b Q_2+\varphi)-k (Q_2-Q_1)+\xi_2,\label{lang2a}
\end{eqnarray}

The commutators and anti-commutators of the quantum noise operators are   given by
\begin{eqnarray}
&&[\xi_i(t),\xi_j(t')]= \delta_{ij}\int\frac{d\omega}{2\pi} N_i(\omega)\hbar\omega\E^{-\ii \omega(t-t')},\label{acom}
\\
&&\average{\{ \xi_i(t),\xi_j(t')   \}}=
\delta_{ij}\int\frac{d\omega}{2\pi}N_i(\omega)\hbar\omega\E^{-\ii \omega(t-t')}\coth\frac{\hbar\omega}{2T_i}.\nonumber \\
\label{aanticom}
\end{eqnarray}
with two time correlation
\begin{eqnarray}
&&\langle\xi_i(t)\xi_j(t')\rangle=\delta_{ij}\int\frac{d\omega}{2\pi}N_i(\omega)\frac{\hbar\omega}2  \E^{-\ii \omega(t-t')} \p{1+ \coth\frac{\hbar\omega}{2T_i}}.\nonumber \\
\label{qcorr1}
\end{eqnarray}
which in Fourier space reads
\begin{eqnarray}
&&\langle\tilde \xi_i(\omega)\tilde \xi_j(\omega')\rangle=\delta_{ij}2 \pi\delta(\omega+\omega') N_i(\omega) \tilde F_i(\omega),
\label{qcorr1F}
\end{eqnarray}
and where we have introduced
\begin{eqnarray}
\tilde F_i(\omega)=\frac {\hbar\omega}{2}   \p{1+ \coth\frac{\hbar\omega}{2T_i}}.\label{def:F}
\end{eqnarray}
In the previous equations we have taken $k_B=1$, a simplification that we will keep in the following.

In order to simplify the notation in the following we will introduce the dimensionless space coordinates $q_i=bQ_i$.

Furthermore we will assume that  the density of states for the baths are equal $N_1(\omega)=N_2(\omega)=N(\omega)$, and thus $\eta_1(t)=\eta_2(t)=\eta(t)$.
In principle $N(\omega)$ is non-zero for $\omega\ge0$, but in order  to make the integrals in the following sections symmetric we can take $N(\omega)=N(-\omega)$, with $\omega$ ranging over the entire real axis.
The density of states takes the form 
\begin{equation}
N(\omega)=2 \eta_0 f(\omega),
\label{N:def}
\end{equation} 
where typical choices for the cutoff function $f(\omega)$ are
 $f(\omega)=1$ (Ohmic bath), or with a soft cutoff $f(\omega)=\Lambda^2/(\omega^2+\Lambda^2)$ or $f(\omega)=\exp(-|\omega|/\Lambda)$.
\section{perturbation expansion}
\label{pert:sec}

A simple symmetry argument indicates that, if the system exhibits a non-vanishing velocity, then it is invariant under sign inversion of the undulation amplitude $V_0\to -V_0$. Indeed inspection of the Hamiltonian (\ref{H0:eq}) suggests that changing sign to $V_0$ corresponds to a translation of the coordinates $Q_i\to Q_i\pm \pi/b$ which does not change the  phase shift between the potentials $V_1$ and $V_2$ in eq.~(\ref{V:eq}). \alb{In other words, the Heisenberg equations (\ref{lang1a})--(\ref{lang2a}) would be the same for the shifted coordinates under sign inversion of $V_0$.}
Thus we conclude that any systematic non-vanishing velocity must be even in $V_0$.

Therefore, following \cite{Aslangul87} we expand the dimensionless coordinates $q_1,\, q_2$ up to second order in the particle potential amplitude $V_0$,
\begin{equation}
q_i=\qo_i+V_0 \qu_i+V_0^2 \qd_i + O(V_0^3).
\end{equation} 
To zeroth order the quantum Langevin equations~(\ref{aqlan}) read
\begin{eqnarray}
m \ddot q_1^{(0)}&=&-\int^t_{t_0} \eta(t-t')\dot q^{(0)}_1(t') \D t'-k (\qo_1-\qo_2) +b \xi_1, \nonumber \\
&&\\
m \ddot q_2^{(0)}&=&-\int^t_{t_0} \eta(t-t')\dot q^{(0)}_2(t') \D t'-k (\qo_2-\qo_1) +b \xi_2.\nonumber \\
&&
\end{eqnarray} 
It is furthermore convenient to introduce the coordinates
\begin{equation}
x(t)=q_1+q_2,\qquad y(t)=q_1-q_2,
\end{equation} 
and their corresponding power series.
To the zero-th order, the coordinate $\Xo(t)$ describes the motion of a free Brownian particle, while $\Yo(t)$ describes a particle in a harmonic potential with strength $2 k$, and their quantum Langevin equations read
\begin{eqnarray}
m \ddot x^{(0)}&=&-\int^t_{t_0} \eta(t-t')\dot x^{(0)}_1(t') \D t' +b (\xi_1+\xi_2),\nonumber \\
m \ddot y_2^{(0)}&=&-\int^t_{t_0} \eta(t-t')\dot y^{(0)}_2(t') \D t'-2 k y^{(0)} +b( \xi_1-\xi_2).\nonumber 
\end{eqnarray} 
Since we are interested in the long-time behaviour of the system,  we will set $t_0=-\infty$ so as the one time averages will be time independent, and the two-time correlation functions will be time translationally invariant.
To zero-th order we have thus that the Green's function solutions of eq.~(\ref{aqlan})  in terms of the coordinate $x$ and $y$ read
\begin{eqnarray}
\Xo(t)&=&b \int_{-\infty}^{t} G_x(t-t') \pq{\xi_1(t')+\xi_2(t')} \D t',\label{X0}\\
\Yo(t)&=&b \int_{-\infty}^{t} G_y(t-t') \pq{\xi_1(t')-\xi_2(t')} \D t',\label{Y0}
\end{eqnarray} 
where the Fourier transform of $G_x(\tau)$ and $G_x(\tau)$ read
\begin{eqnarray}
\tilde  G_x(\omega)&=& -(m \omega^2+ \ii \omega \tilde \eta(\omega))^{-1},\label{Gx:def}\\
\tilde  G_y(\omega)&=& -(m \omega^2+ \ii \omega \tilde \eta(\omega)- 2 k )^{-1}. \label{Gy:def}
\end{eqnarray} 
Thus one finally obtains the zeroth-order solutions
\begin{eqnarray}
\qo_1&=&(\Xo+\Yo)/2, \label{q10}\\
\qo_2&=&(\Xo-\Yo)/2.\label{q20}
\end{eqnarray}

It is straightforward to show that the commutators $\pq{\qo_i(t_1),\qo_j(t_2)}$ are complex-valued functions of the time lapse $t_1-t_2$.
We have indeed
\begin{eqnarray}
\tilde q_1^{(0)}(\omega)&=&\frac b 2 \left[ (\tilde G_x(\omega) +  \tilde G_y(\omega)) \tilde \xi_1(\omega)\right.\nonumber \\
&&\quad \left .+  (\tilde G_x(\omega) -  \tilde G_y(\omega)) \tilde \xi_2(\omega)\right] \label{q1om}\\
\tilde q_2^{(0)}(\omega)&=&\frac b 2 \left[ (\tilde G_x(\omega) -  \tilde G_y(\omega)) \tilde \xi_1(\omega)\right.\nonumber \\
&&\quad \left .+  (\tilde G_x(\omega) +  \tilde G_y(\omega)) \tilde \xi_2(\omega)\right] \label{q2om}
\end{eqnarray} 
By taking into account that $\tilde G_{x,y}(-\omega)=\tilde G^*_{x,y}(\omega)$, and by using eq.~(\ref{acom}), 
we can then calculate 
\begin{eqnarray}
&&\pq{\tilde q_1^{(0)}(\omega),\tilde q_1^{(0)}(\omega')}=\pq{\tilde q_2^{(0)}(\omega),\tilde q_2^{(0)}(\omega')}\nonumber\\
&&=b^2 \pi  N(\omega) \hbar \omega  \delta(\omega+\omega')  \times \p{|\tilde G_x(\omega)|^2 +|\tilde G_y(\omega)|^2}\\
&&\pq{\tilde q_1^{(0)}(\omega),\tilde q_2^{(0)}(\omega')}=\nonumber\\
&&=b^2 \pi  N(\omega) \hbar \omega  \delta(\omega+\omega')  \times \p{|\tilde G_x(\omega)|^2 -|\tilde G_y(\omega)|^2}
\end{eqnarray} 
obtaining thus
\begin{eqnarray}
&&\pq{\qo_1(t),\qo_1(t')}=\pq{\qo_2(t),\qo_2(t')}=\nonumber\\
&&=b^2 \int \frac{\D \omega}{2 \pi }N(\omega) \frac{\hbar \omega}{2}  \p{|\tilde G_x(\omega)|^2 +|\tilde G_y(\omega)|^2} \E^{-\ii \omega(t-t')}\label{corr11}\\
&&\pq{\qo_1(t),\qo_2(t')}=\nonumber\\
&&=b^2 \int \frac{\D \omega}{2 \pi }N(\omega) \frac{\hbar \omega}{2}  \p{|\tilde G_x(\omega)|^2 -|\tilde G_y(\omega)|^2} \E^{-\ii \omega(t-t')}. \label{corr12}
\end{eqnarray} 
In order to shorten the notation in the following we set
\begin{equation}
a_{ij}(t-t')\equiv \frac {\ii}{2} \pq{\qo_i(t),\qo_j(t')}.
\label{def:aij}
\end{equation} 
and 
\begin{equation}
A_{ij}(t-t')\equiv a_{ij}(t-t')/b^2=\frac {\ii}{2} \pq{\Qo_i(t),\Qo_j(t')},
\label{def:Aij}
\end{equation} 
which are $c$-numbers.
We notice for later use that $a_{12}(t-t')=a_{21}(t-t')$, which can be obtained from eq.~(\ref{corr12}) by taking into account that $N(\omega),\, |\tilde G_x(\omega)|^2$ and $|\tilde G_y(\omega)|^2$ are all even functions of $\omega$.

Similarly, from eqs.~(\ref{qcorr1F}) and (\ref{q1om})-(\ref{q2om})  we can calculate the correlations $\average{\qo_i(t)\qo_j(t')}$ (see appendix \ref{app1})
and from them one can obtain the quantities
\begin{widetext}
\begin{eqnarray}
c_{12}(t-t')&=&\average{(\qo_1(t)-\qo_2(t'))^2}=\nonumber\\
&=&\frac{b^2} 2 \int \frac{\D \omega}{2 \pi}N(\omega)\left[\tilde F_1(\omega)+ \tilde F_2(\omega)\right] \pq{|\tilde G_x(\omega)|^2 (1-\cos \omega(t-t')) +|\tilde G_y(\omega)|^2 (1+\cos \omega(t-t'))}\nonumber \\
&& \qquad \quad - \ii  N(\omega)\left[\tilde F_1(\omega)- \tilde F_2(\omega)\right]  \sin \omega(t-t') (\tilde G_x(\omega)\tilde G_y(-\omega)-\tilde G_x(-\omega)\tilde G_y(\omega))
\label{C12}
\end{eqnarray} 
\end{widetext}
Similarly one finds 
\begin{equation}
c_{21}(t-t')=\langle(\qo_2(t)-\qo_1(t'))^2\rangle=c_{12}(t'-t)
\label{C21}
\end{equation} 

\subsection{First order}
In order to simplify the calculations in the following we will take the phase appearing in eq.~(\ref{V:sin}) to be $\varphi=\pi/2$. We will then generalise our results to the case of arbitrary $\varphi$.
To first order in $V_0$,  eqs.~(\ref{aqlan}) become 
\begin{eqnarray}
M\ddot q^{(1)}_1&=&-\int_{-\infty}^t \eta(t-t') \dot q^{(1)}_1(t') \D t' \nonumber \\
&&- b^2 \sin \qo_1-k (\qu_1-\qu_2),\label{lang11}\\
M \ddot q^{(1)}_2 &=&-\int_{-\infty}^t \eta(t-t') \dot q^{(1)}_2(t') \D t' \nonumber \\
&&- b^2 \cos \qo_2-k (\qu_2-\qu_1),
\label{lang21}
\end{eqnarray} 
with solutions 
\begin{eqnarray}
\qu_1&=&(\Xu+\Yu)/2, \label{q11}\\
\qu_2&=&(\Xu-\Yu)/2,\label{q21}
\end{eqnarray} 
where
\begin{eqnarray}
\Xu(t)&=&-b^2\int_{-\infty}^{t} G_x(t-t') \pq{\sin \qo_1(t')+\cos \qo_2(t')} \D t',\nonumber \\
&&\label{X1}\\
\Yu(t)&=&-b^2\int_{-\infty}^{t} G_y(t-t') \pq{\sin \qo_1(t')-\cos \qo_2(t')} \D t' .\nonumber \\
&& \label{Y1}
\end{eqnarray} 

We can now check that the first order term in $V_0$ of the velocity vanishes.
One possible way to proceed is to differentiate eq.~(\ref{X1}), and to perform a statistical average over the quantum noise operators $\xi_i(t)$. 
One thus obtains 
\begin{eqnarray}
\average{\dot x^{(1)}}&=& b^2 \int_{-\infty}^{t} \partial_{t'}G_x(t-t') \average{\sin \qo_1(t')+\cos \qo_2(t')} \D t' \nonumber \\
&&-b^2 G_x(0) \average{\sin \qo_1(t)+\cos \qo_2(t)}.
\label{eqx1d}
\end{eqnarray} 
The calculation of the averages can be performed by expressing the trigonometric functions in their exponential forms. Thus one has to evaluate terms of the form $\average{\exp(\pm\ii q_i(t))}$. This can be done by noticing that the position operators $\qo_i(t)$ are linear combinations of the noise operators $\xi_i$, eqs.~(\ref{X0})--(\ref{Y0}), which in turn are normally distributed with zero mean. 
If $X$ is a normally distributed variable, with zero mean, one easily finds that $\average{\exp (X)}=\exp( \average{X^2}/2$). By applying this equality to the trigonometric functions on the right hand side (rhs) of eq.~(\ref{eqx1d}), one finds  $\average{\sin \qo_1(t)}=0$, and given that $\average{(\qo_2(t))^2}$ is independent of the time, (see appendix \ref{app1}), also $\average{\cos \qo_2(t)}$ it time independent. Thus we conclude that the rhs of eq.~(\ref{eqx1d}) vanishes.
\subsection{Second order}
To second order in $V_0$, eqs.~(\ref{lang1a})--(\ref{lang2a}) become 
\begin{eqnarray}
M\ddot q^{(2)}_1&=&-\int_{-\infty}^t \eta(t-t') \dot q^{(2)}_1(t') \D t'-k (\qd_1-\qd_2)\nonumber\\
&& - b^2 \pg{\sin (\qo_1+ V_0\qu_1)}^{(1)},\label{lang12}\\
M \ddot q^{(2)}_2 &=&-\int_{-\infty}^t \eta(t-t') \dot q^{(2)}_2(t') \D t'-k (\qd_2-\qd_1) \nonumber\\
&& -b^2 \pg{\cos (\qo_2+ V_0\qu_2)}^{(1)},
\label{lang22}
\end{eqnarray} 
where the notation $\pg{\sin (\qo_1+ V_0\qu_1)}^{(1)}$ and $\pg{\cos (\qo_2+ V_0\qu_2)}^{(1)}$ indicates the first order term in the expansion. However 
the expansion of the trigonometric functions must be taken carefully, given that the operators $\qo_i(t)$ and $\qu_i(t)$ do not commute.

Assuming that the system reaches a constant average center-of-mass velocity in the steady state (which can well be vanishing) is equivalent to take $\average{\ddot  x^{(2)}}=0$, where the average is taken over the quantum noise operators $\xi_i$.  Thus summing up eqs.(\ref{lang12})-(\ref{lang22}), and averaging over the quantum noise,  we obtain the following equation for $\dot x^{(2)}(t)$ 
\begin{eqnarray}
&&\int^t_{t_0} \eta(t-t')\langle \dot x^{(2)}(t')\rangle \D t'=\nonumber\\
&&=-b^2 \average{\pg{\sin (\qo_1+ V_0\qu_1)}^{(1)}+\pg{\cos (\qo_2+ V_0\qu_2)}^{(1)}} \nonumber\\
\label{aveX}
\end{eqnarray} 

The next step is  to expand the terms on the right hand side of the last equation up to the first order in $V_0$.
\alb{The derivation is long but quite straightforward, and is presented in appendix~\ref{app1a}.}

\alb{For the second order contribution to the steady velocity one thus obtains}
\begin{equation}
 \langle\dot x^{(2)}\rangle=b^2 \frac{\mathcal I}{\tilde \eta(0)},
\label{avex2}
\end{equation} 
where
\begin{eqnarray} 
&&\mathcal I=\frac 1 4  \int_{0}^{+\infty} \D\tau \, (G_x(\tau)-G_y(\tau)) \nonumber \\
&&\qquad \qquad \frac{\sin a_{12}(\tau)}{ A_{12}(\tau)}\p{\E^{-\frac 1 2 c_{12}(\tau)}-\E^{-\frac 1 2 c_{21}(\tau)}}
\label{defI}
\end{eqnarray}

Thus, up to second order in $V_0$ the velocity of the center of mass reads
 \begin{equation}
\bar v^{(2)}=\frac 1 2 \langle \dot Q_1^{(2)} + \dot Q_2^{(2)} \rangle= V_0^2 b \frac{\mathcal I}{2 \tilde \eta(0)}+O(V_0^4).
\label{v02}
\end{equation} 

We now consider the case where the potentials $V_i$ have  an arbitrary phase shift $\vphi$: $V_2(Q_2)=-V_0 \cos(b Q_2+\varphi)$.
By introducing the shifted variable $\bar q^{(0)}_2=\qo_2+\vphi$ and retracing the previous steps, where we have taken the specific value $\vphi=\pi/2$,  one obtains for the steady state velocity up to the second order in $V_0$ 
 \begin{equation}
\bar v^{(2)}=\frac 1 2 \langle \dot Q_1 + \dot Q_2  \rangle= V_0^2 b \sin \varphi  \frac{\mathcal I}{2 \tilde \eta(0)}+O(V_0^4).
\label{v2b}
\end{equation} 

 \alb{ In this section we have derived the result~(\ref{avex2})  by assuming that $\average{\ddot  x^{(2)}}=0$. While it is reasonable to assume that in the steady state the average center-of-mass velocity is constant, in appendix \ref{app1ax} we provide an exact proof of the result  $\average{\ddot  x^{(2)}}=0$}.

\section{Second order velocity term: analysis}
\label{sec:analysis}
\begin{figure}[h]
\center
\includegraphics[width=8cm]{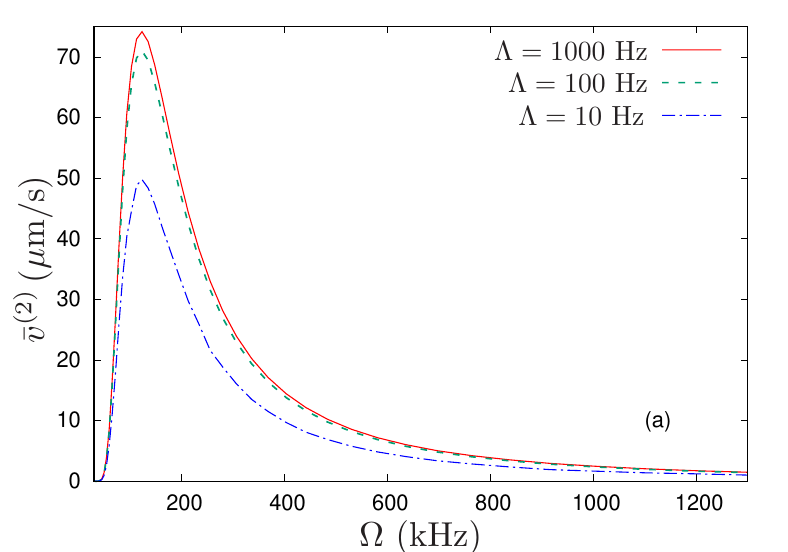}
\includegraphics[width=8cm]{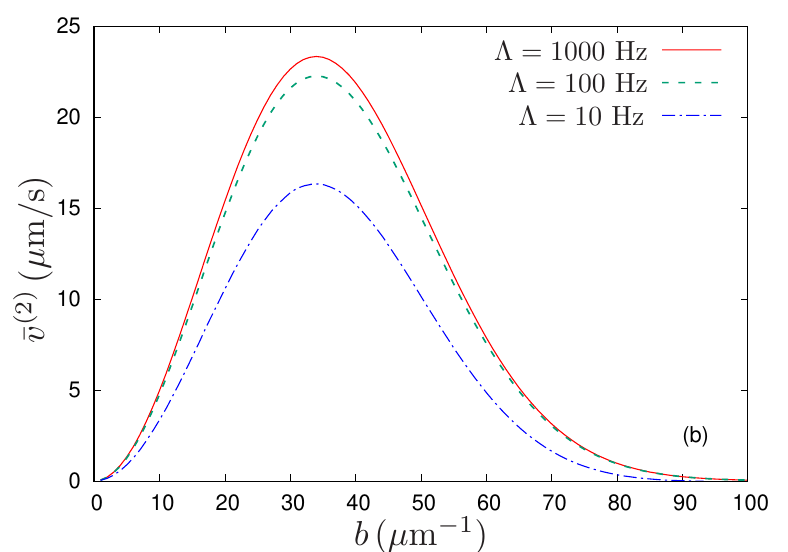}
\includegraphics[width=8cm]{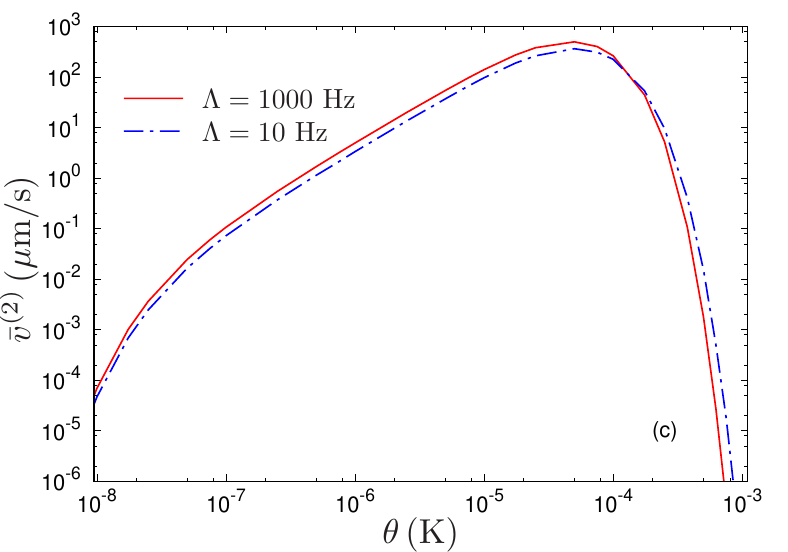}
\includegraphics[width=8cm]{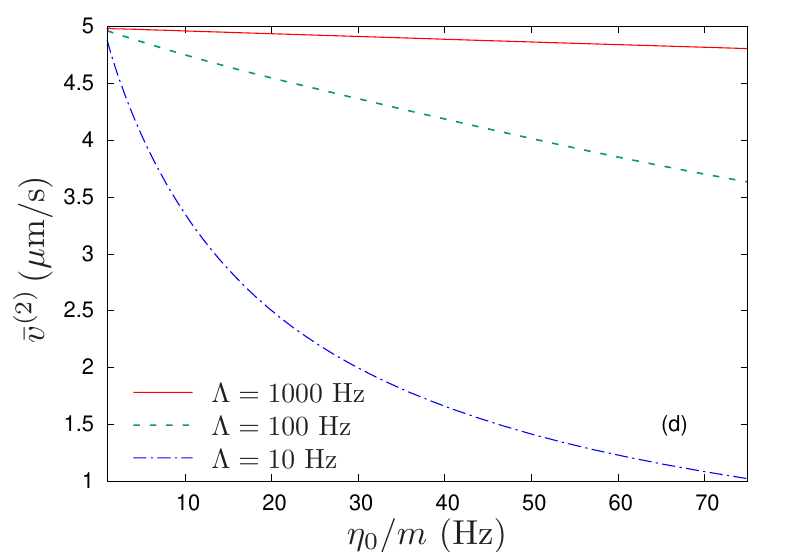}
\caption{
 Steady state center-of-mass velocity $\bar v^{(2)}$ as given by eq.~(\ref{v02}), for different cutoff frequencies.
Here we use the soft cutoff function $f(\omega)=\Lambda^2/(\omega^2+\Lambda^2)$ for the density of states. 
Panel (a): $\bar v^{(2)}$  as a function of the  particle-particle interaction frequency $\Omega=\sqrt{k/m}$
 , with $b=10\, \mu\mathrm{m}^{-1}$, $T_1= 1 \mu K$, $T_2=2.5 T_1$,  $\eta_0/m=10$ Hz. 
Panel (b):  $\bar v^{(2)}$ as a function of the  wavenumber $b$,  with $\Omega=\sqrt{k/m}=702.5$ kHz, $T_1= 1 \mu K$, $T_2=2.5 T_1$, $\eta_0/m=10$ Hz.
Panel (c):  $\bar v^{(2)}$ as a function of the temperature scale $\theta$, with $T_1= \theta $, $T_2=2.5 \theta$, $b=10\, \mu\mathrm{m}^{-1}$, $\Omega=\sqrt{k/m}=702.5$ kHz, $\eta_0/m=10$ Hz.
Panel (d):   $\bar v^{(2)}$ as a function of the friction coefficient $\eta_0$,  with $b=10\, \mu\mathrm{m}^{-1}$, $T_1= 1 \mu K$, $T_2=2.5 T_1$, $\Omega=\sqrt{k/m}=702.5$ kHz.
 The values of the other parameters are  $m=40$ amu, $\vphi=\pi/2$,  $V_0=T_1/4$.
}
\label{figb}
\end{figure}

We now analyze the results for the constant second order velocity, eqs.~(\ref{avex2})--(\ref{v2b}).
Plots of the system center of mass velocity as a function of the interaction strength $k$, of the wavenumber $b$, and of the temperature scale are shown in fig.~\ref{figb} for the soft cutoff function $f(\omega)=\Lambda^2/(\omega^2+\Lambda^2)$. 
The physical parameter space considered in Fig. 1 has been chosen in order to analyze a potential experimental realization of the model where the two particles are represented by two sideband laser-cooled atomic ions.  Confinement of the ions by a common harmonic trap potential will lead to an effective harmonic binding force between them \cite{Grimm2000}, while spatially periodical potentials along the axis defined by the two ions can be realized through off-resonant electrical dipole forces induced by standing wave light fields \cite{Linnet12,Enderlein2012,Bylinskii2015,Laupretre2019}. By either choosing two identical ion species initialized in different internal states or two different ion isotopes, different polarization states and/or longitudinal modes of an Fabry-Perot cavity could enable particle dependent periodical potentials with a constant, but tunable phase relation between them \cite{Linnet2014}.

When does the right hand side term of equation~(\ref{defI}), and thus $\bar v^{(2)}$ vanish?
From eq.~(\ref{def:F}) we see that $\tilde F_1(\omega)= \tilde F_2(\omega)$ when $T_1=T_2$. By inspecting eqs.~(\ref{C12})--(\ref{C21}) we find that $c_{12}(t-t')=c_{21}(t-t')$ for equal temperatures. Therefore we conclude that the velocity  $\bar v^{(2)}$ vanishes at thermal equilibrium, as expected: it is the heat current flowing between the two baths that sustains a non-vanishing velocity.

Furthermore the the center-of-mass velocity vanishes in the trivial case of vanishing undulation amplitude $V_0=0$, i.e., the motor requires {\it gear racks} in order to work, which are represented by the two periodic potentials $V_1$ and $V_2$.

Most importantly,  we notice that the second order term of the velocity vanishes for a phase shift between the potentials $\varphi=l \pi$ with $l$ an integer number.
This is consistent with the results for the classical counterpart of the present model found in \cite{Fogedby17}, where it was numerically shown that the velocity vanishes for  $\varphi=l \pi$  independently of the order of $V_0$.
In this case the system does not break the spatial symmetry discussed in section \ref{model:sec}. Such a broken symmetry has been shown to be  a prerequisite  in order for directed motion in classical and quantum duets to arise \cite{Fogedby17,Hovhannisyan_2019}, as well as in many-body systems \cite{Sune19a} .
 We recall that the required broken symmetry reads as follows: there is no  translation distance $\Delta$ such that   $V(-Q_1,-Q_2)=V(Q_1+\Delta,Q_2)$. 
 Such a broken symmetry is the 2D counterpart of the broken symmetry discussed in \cite{Reimann02} for 1D non--autonomous  Brownian motors. In 1D this broken spatial symmetry amounts to require that the potential is, e.g., saw-tooth shaped.

One should also expect that the velocity vanishes in the limit $k\to 0$ (decoupled system) and in the limit of tight coupling $k\to \infty$. In the latter case the system behaves as a single particle in contact with two environments at different temperatures, for which the fluctuations of the relative coordinate are suppressed $\average{y}\to 0$, leading to $\tilde G_y(\omega)\to 0$, which in turn implies that the asymmetric term in (\ref{C12}) vanishes. 
\alb{In simpler terms, in the limit $k\to \infty$ the sinusoidal tracks, which are responsible for the broken spatial symmetry and thus for the motor effect, become negligible with respect to the interaction potential~(\ref{U:quad}), as $V_0/k \to 0$.}

Therefore one should expect an optimal coupling strength between these two regimes, as confirmed by inspection of fig.~\ref{figb}-(a), where we plot the second order steady state velocity $\bar v^{(2)}$ as a function of the particle-particle interaction strength for different values of the cutoff frequency.
Furthermore from eq.~(\ref{v2b})  we find that the velocity, up to the second order in $V_0$, vanishes in the limit of $b\to 0$ (infinite period). Inspection of eqs.~(\ref{lang1a})--(\ref{lang2a}) suggests that  the center of mass velocity must also vanish  in the limit $b \to \infty$ as the force exerted by the particles' potentials will prevail on the bath forces $\xi_i$ which  drive the directed motion.
Thus one must expect an optimal value of $b$ for which the velocity achieves a maximum.
This is confirmed by inspection of fig.~\ref{figb}-(b) where we plot $\bar v^{(2)}$ as a function of the wavenumber $b$,  for different values of the cutoff frequency. 
\onecolumngrid

\twocolumngrid

The velocity $\bar v^{(2)} $ is expected to vanish in both the large and the low temperature regime. The limit of large $T_1$ and $T_2$ corresponds trivially to the limit $V_0\to 0$, where the gears represented by the periodic tracks are flattened. In the limit of low temperatures it is the heat currents that fuels the motor that vanishes. These considerations are confirmed by inspection of fig.~(\ref{figb})-(c), where the second order steady state velocity is plotted as a function of the temperature scale.


\alb{The friction coefficient appearing in the expression of the density of states  $N(\omega)$, eq.~(\ref{N:def}), plays the role of an effective coupling strength with the baths' oscillators, see also eq.~(\ref{aDOS}). It is thus interesting to evaluate the velocity $\bar v^{(2)} $ as a function of such a parameter. One expects that for large $\eta_0$, the motor is so strongly coupled with the baths' degrees of freedom, that its velocity should vanish. Similarly for vanishing $\eta_0$, the motor velocity should vanish, as the propelling effect of the baths would also vanish. However, one should keep in mind that for very small values of $\eta_0$ the integrand in eq.~(\ref{defI}) becomes highly oscillating and its numerical evaluation becomes unfeasible. We therefore consider here the lower bond $\eta_0/m >1$ Hz. The results for the velocity as function of $\eta_0$ are shown in  fig.~(\ref{figb})-(d), and we find that in the range $\eta_0/m >1$ Hz the velocity decreases with different rates for increasing  $\eta_0$, depending on the cutoff frequency $\Lambda$.}

Inspections of the four panels in fig.~\ref{figb} indicates that in general the velocity increases for larger cutoff $\Lambda$. This can be easily understood by noticing that it is the modes in the baths with density of states $N(\omega)$ as given by eq.~(\ref{N:def}) that propel the motor. The larger the cutoff $\Lambda$ the more modes are present in the bath, with higher frequency, and thus larger average thermal energy.
This is no longer true in the limit of larger temperatures, see  fig.~\ref{figb}-(c). This can be understood by considering that, as discussed above, the limit of large temperatures correspond to the limit of $V_0\to 0$. So having fewer modes in the baths (with lower average energy) is beneficial for the motor as the flattening of the tracks is reduced with respect to the case of large $\Lambda$.
It is important to notice that the results obtained in this section  for  $\bar v^{(2)} $ are exact up to the second order in $V_0$. Thus $V_0$ has to be small compared to the other energy scales in the system: $k_B T_i$, $k/b^2$ and $\hbar \sqrt{k/m}$.
In this regard, if one considers the limit where one of the two temperatures vanishes, e.g. $T_1\to 0$, while the second is finite $T_2>0$, the potential amplitude must also go to zero $V_0\to 0$, and in this limit the motor effect vanishes, as discussed above.  

The above results for the velocity, and in particular eqs.~(\ref{defI})--(\ref{v02}),  hold true also in the classical limit of large temperatures ($\hbar \to 0$), provided that in the expression of the correlation  $\langle ( \qo_1(\tau) - \qo_2(t))^2\rangle$  eq.~(\ref{C12}), one takes  $\tilde F_i(\omega)\to T_i$ for  $\hbar \to 0$, and by noticing that
 from eq.~(\ref{def:aij}) one finds  
\begin{equation}
\lim_{\hbar\to0} \sin(a_{ij}(\tau))/a_{ij}(\tau)=1.
\end{equation} 

\section{External forces}
\label{sec:F}
We now consider the case where two external forces $F_i(t)$ are applied on the two particles. In terms of the rescaled coordinates the quantum Langevin equations~(\ref{aqlan})  become
\begin{eqnarray}
M\ddot q_1&=&-\int^t_{-\infty} \eta(t-t') \dot q_1(t') \D t' - b^2 V_0 \sin q_1-k (q_1-q_2) \nonumber\\
&&+ b \xi_1+ b F_1, \label{lang1c}\\
M \ddot q_2 &=&-\int^t_{-\infty} \eta(t-t') \dot q_2(t') \D t' - b^2 V_0 \sin( q_2+ \varphi)\nonumber\\
&&-k (q_2-q_1)+ b \xi_2+ b F_2.
\label{lang2c}
\end{eqnarray}
 
To zeroth order the solutions of eqs.~(\ref{lang1c})--(\ref{lang2c}) consist now of two contributions, one arising from the random forces $\xi_i(t)$  and
one from the systematic forces $F_i(t)$, and read 
\begin{eqnarray}
\Xo(t)&=&\Xo_R(t)+\Xo_S(t)\nonumber \\
&=&b \int_{-\infty}^{t} G_x(t-t')\nonumber \\ 
&& \quad \times \pq{\xi_1(t')+\xi_2(t')+F_1(t')+F_2(t')} \D t',\label{X0F}
\end{eqnarray} 
\begin{eqnarray}
\Yo(t)&=&\Yo_R(t)+\Yo_S(t)\nonumber \\
&=&b \int_{-\infty}^{t} G_y(t-t')\nonumber \\ 
&& \quad \times \pq{\xi_1(t')-\xi_2(t')+F_1(t')-F_2(t')} \D t',\label{Y0F}
\end{eqnarray} 
with $\qo_{1,2}=(\Xo\pm\Yo)/2$.
By comparing the equations eqs.~(\ref{X0F})-(\ref{Y0F}) with eqs.~(\ref{X0})-(\ref{Y0}) one finds that, as expected, the random part of the solutions $\Xo_R(t)$, $\Yo_R(t)$ are identical to the solutions  $\Xo(t)$, $\Yo(t)$ discussed in sec.~\ref{pert:sec}.
Similarly, to the first order,  one finds  
\begin{eqnarray}
\Xu(t)&=&-b^2\int_{-\infty}^{t} G_x(t-t') \pq{\sin \qo_1(t')+\cos \bar q^{(0)}_2  (t')} \D t',\nonumber \\
&&\label{X1F}\\
\Yu(t)&=&-b^2\int_{-\infty}^{t} G_y(t-t') \pq{\sin \qo_1(t')-\cos \bar q^{(0)}_2 (t')} \D t' \nonumber \\
&& \label{Y1F}
\end{eqnarray} 
which are identical to the first order  solutions (\ref{X1})--(\ref{Y1}) in sec.~\ref{pert:sec}.

Finally, to second order in $V_0$ one obtains
\begin{eqnarray} 
 &&M\ddot q^{(2)}_1+\int_{-\infty}^{t} \eta(t') \dot q^{(2)}_1(t')\D t'+k (q^{(2)}_1-q^{(2)}_2)\nonumber \\
&& =- b^2 \pg{\sin  (\qo_1 + V_0 q^{(1)}_1)}^{(1)},\label{lang1d}\\
&&M\ddot q^{(2)}_2+\int_{-\infty}^{t} \eta(t') \dot q^{(2)}_2(t')\D t'+k (q^{(2)}_2-q^{(2)}_1)\nonumber \\
&& =- b^2 \pg{\sin  (\bar q^{(0)}_2 + V_0 q^{(1)}_2)}^{(1)}\label{lang2d},
\end{eqnarray} 
which are identical to the second order quantum Langevin equations~(\ref{lang12})--(\ref{lang22}) in absence of external force.
Thus, up to this point, the only difference when one applies the external force $F_i(t)$ is in the zeroth order solutions (\ref{X0F})--(\ref{Y0F}).
In particular the calculation of the first order expansion of the rhs of equations~(\ref{lang1d})--(\ref{lang2d}) can be carried out as in appendix~\ref{app1a} obtaining identical expressions to eqs.~(\ref{sinexp})--(\ref{cosexp}).
As in section ~\ref{pert:sec} and in appendix~\ref{app1a}  we need to calculate the average of the rhs of eqs.~(\ref{lang1d})--(\ref{lang2d}) over the bath variables $\xi_i(t)$.
However, when averaging the rhs of  eqs.~(\ref{sinexp})--(\ref{cosexp}), we have to take into account that the averages  $\langle \sin \qo_i(\tau) \cos \qo_i(t)\rangle$ do not vanish, as the zeroth order solution contains now a contribution from the deterministic forces $F_i(t)$, see eqs.~(\ref{X0F})--(\ref{Y0F}).
With these considerations in mind we can thus calculate the average of  the rhs of~(\ref{lang1d})--(\ref{lang2d}) and obtain
\begin{widetext}
\begin{eqnarray} 
&&\langle M\ddot q^{(2)}_1+\eta \dot q^{(2)}_1+k (q^{(2)}_1-q^{(2)}_2)\rangle=\nonumber \\
&&\frac{b^2}{4 }\int_{-\infty}^t \D \tau [G_x(t-\tau)+G_y(t-\tau)]\left\{ \frac{\sin a_{11}(\tau-t)}{a_{11}(\tau-t)}\E^{-\frac 1 2 c_{11}(\tau-t)}\sin( \qo_{1,S}(\tau) - \qo_{1,S}(t))\right\}\nonumber\\
 &&\qquad \quad+ [G_x(t-\tau)-G_y(t-\tau)] \left\{\frac{\sin a_{21}(\tau-t)}{ a_{21}(\tau-t)}\E^{-\frac 1 2 c_{21}(\tau-t)}\sin( \qo_{2,S}(\tau) - \qo_{1,S}(t)+\vphi) \right\}\label{q12F}\\
&&\langle M\ddot q^{(2)}_2+\eta \dot q^{(2)}_2+k (q^{(2)}_2-q^{(2)}_1)\rangle=\nonumber\\
&&\frac{b^2}{4 }\int_{-\infty}^t \D \tau [G_x(t-\tau)+G_y(t-\tau)]\left\{ \frac{\sin a_{22}(\tau-t)}{a_{22}(\tau-t)}\E^{-\frac 1 2 c_{22}(\tau-t)}\sin( \qo_{2,S}(\tau) - \qo_{2,S}(t))\right\}\nonumber\\
 &&\qquad \quad+ [G_x(t-\tau)-G_y(t-\tau)] \left\{\frac{\sin a_{12}(\tau-t)}{ a_{12}(\tau-t)}\E^{-\frac 1 2 c_{12}(\tau-t)}\sin( \qo_{1,S}(\tau) - \qo_{2,S}(t)-\vphi) \right\}\label{q22F}
\end{eqnarray} 
\end{widetext}
\subsection{Constant forces}
\label{subsec:F}
We now consider the case in which the deterministic forces are constant $F_i(t)=\bar F_i$.  
The system will thus reach a steady state with a constant average velocity.
By summing  eqs.~(\ref{q12F})--(\ref{q22F}), and exploiting the fact that the two time commutators for $\qo_1$ and $\qo_2$ are identical (eq.~(\ref{corr11})), we obtain for the center of mass 
\begin{equation}
\int_{-\infty}^t \eta(t-t')\langle  \dot x ^{(2)}(t')\rangle\D t'=b^2 \mathcal{I}_F
\end{equation} 
where
\begin{widetext}
\begin{eqnarray} 
\mathcal{I}_F&&=\frac{1}{8 }\int_{-\infty}^t \D \tau [G_x(t-\tau)+G_y(t-\tau)] \frac{\sin a_{11}(\tau-t)}{A_{11}(\tau-t)}\left\{\E^{-\frac 1 2 c_{11}(\tau-t)}\sin( \qo_{1,S}(\tau) - \qo_{1,S}(t))+\E^{-\frac 1 2 c_{22}(\tau-t)}\sin( \qo_{2,S}(\tau) - \qo_{2,S}(t))\right\}\nonumber\\
 &&\qquad\qquad\,  + [G_x(t-\tau)-G_y(t-\tau)] \frac{\sin a_{12}(\tau-t)}{ A_{12}(\tau-t)}\sin( \qo_{2,S}(\tau) - \qo_{1,S}(t)+\vphi)\left\{\E^{-\frac 1 2 c_{21}(\tau-t)}-\E^{-\frac 1 2 c_{12}(\tau-t)} \right\}\label{IF:def},
\end{eqnarray} 
\end{widetext}
with the systematic part of the zeroth order solution given by 
\begin{equation}
q^{(0)}_{1(2),S}(t)=\frac{b}{2}  \int_{-\infty}^t \D t' (\bFu+\bFd) G_x(t-t')\pm (\bFu-\bFd) G_y(t-t'),
\end{equation} 
and 
where the expressions for the correlations $c_{11}(t)$ and $c_{22}(t)$ are given by eq.~(\ref{C11}) in appendix \ref{app1}.

Thus, up to the second order in $V_0$ the velocity of the center of mass reads
 \begin{equation}
\bar v_F^{(2)}= \frac 1 2 \langle \dot Q_1 + \dot Q_2  \rangle \simeq V_0^2 b \frac{\mathcal I_F}{\tilde \eta(0)}+O(V_0^4).
\label{v02F}
\end{equation} 
If the forces $\bar F_{1,2}$ have opposite sign with respect to the system velocity, and are not large enough in modulus to invert the direction of the motion, one can thus extract work from the thermal machine by doing work against such external forces.
If one takes the case $\bFu=\bFd=\bar F$, the output work rate reads
\begin{equation}
\dot W=-\frac {\bar F } 2 \langle \dot Q_1 + \dot Q_2  \rangle = -\bar F V_0^2 b \frac{\mathcal I_F}{\tilde \eta(0)}+O(V_0^4).
\end{equation} 

When one consider the non interacting case $k=0$, the two particles move independently under the effect of the two forces $\bFu$  and $\bFd$. In this limit the  Green's function  $G_y(t)$ becomes  equal to $G_x(t)$ (see eqs. (\ref{Gx:def}) and (\ref{Gy:def})) and the second term in eq.~(\ref{IF:def}) vanishes. This is the term responsible for the thermal propulsion even in absence of external  forces, see eqs.~(\ref{defI})--(\ref{v02}).
On the other hand the first term in eq.~(\ref{IF:def}) is non zero when $\bFu$  or $\bFd\neq 0$.
Thus for $k=0$, eqs.~(\ref{IF:def})--(\ref{v02F}) reduce to the results of  \cite{Fisher85,Aslangul87} for the case of a single quantum Brownian particle in a sinusoidal potential under the effect of a constant external force.

\section{Simulations}

In the Quantum Molecular Dynamics (QMD) algorithm, as introduced in \cite{Dammak09} for systems at equilibrium with a single bath, a quantum thermal bath 
is replaced by a classical bath that accounts for quantum statistics in the framework of a standard MD algorithm.
Specifically, in the QMD algorithm the Heisenberg equations of motion for the quantum operators $Q_i$ and $P_i$
 are replaced by a classical Langevin equation, of the type (\ref{aqlan}),
where the power spectral density of the stochastic noise  is given by the quantum mechanical fluctuation-dissipation relation (\ref{def:F}).
Thus the QMD algorithm neglects one basic quantum feature, namely the  noncommuting character of the system variables, while it retains the power spectral density of the bath variables.

In order to avoid any possible misinterpretation, we remark once more that the formalism in sections~\ref{model:sec}--\ref{sec:F} is purely quantum, as eq.~(\ref{aqlan}) is a dynamical equation for the operators $Q_i$  in
the Heisenberg picture.
Although the  
Ehrenfest theorem prescribes the equivalence of classical and
quantum dynamic equations only for harmonic potentials,  the QMD algorithm has provided accurate results for different types of systems with  various degrees of anharmonicity \cite{Dammak12,Calvo12,Calvo12a,Qi12,Bronstein14}. For example the validity of the approach has been tested by reproducing several equilibrium experimental data at low temperatures in a regime where quantum
statistical effects cannot be neglected, for  MgO crystal or
nonsuperfluid liquid $^4\mathrm{He}$, characterized by anharmonic potentials \cite{Dammak09}.

The QMD has been extended to the non--equilibrium case, with two heat reservoirs at different $T_i$ in ref.~\cite{Fogedby18}, but only for the case of a parabolic potential.

Given that the results contained in  section~\ref{sec:analysis} are exact, up to the second order in $V_0$, the model discussed in the present paper represents an excellent test-bed to check whether the QMD can be used to evaluate the dynamic properties of a system that is both non--linear and out-of-equilibrium. We will consider in the following  Ohmic baths ($\Lambda \to \infty$), while the numerics have been performed by using dimensionless  quantities for the system parameters. \alb{Similarly to ref.~\cite{Dammak09}, we use the numerical method discussed in \cite{Maradudin1990} to generate the correlated noise with fluctuation-dissipation relation (\ref{def:F}).}

\begin{figure}[h]
\center
\psfrag{v}[ct][ct][1.]{$\bar v$}
\psfrag{k}[ct][ct][1.]{$k$}
\psfrag{QMD}[rb][rb][.9]{QMD}
\psfrag{MD}[rb][rb][.9]{MD}
\psfrag{QMDx}[rb][rb][.9]{QMD}
\psfrag{MDx}[rb][rb][.9]{MD}
\psfrag{exact}[rb][rb][.9]{$\bar v^{(2)}, V_0=0.5$}
\psfrag{exact1}[rb][rb][.9]{$\bar v^{(2)}, V_0=0.75$}
\includegraphics[width=8cm]{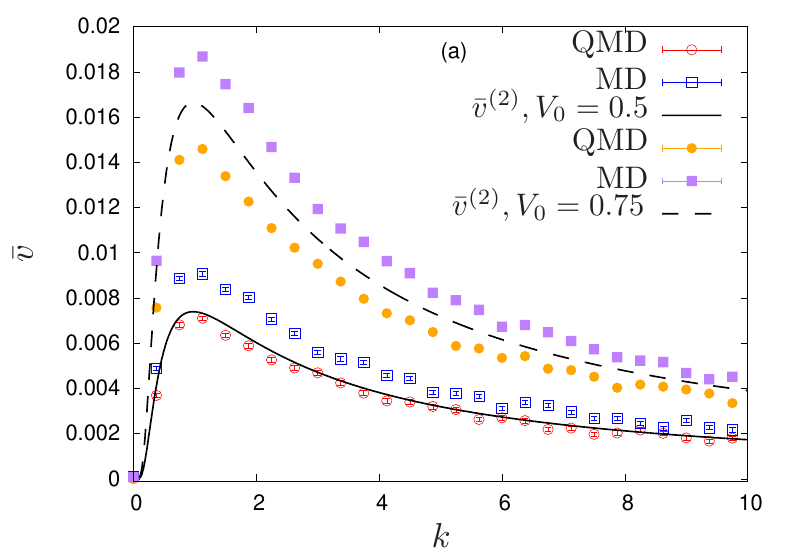}
\psfrag{exact}[rb][rb][.9]{$\bar v^{(2)}$}
\includegraphics[width=8cm]{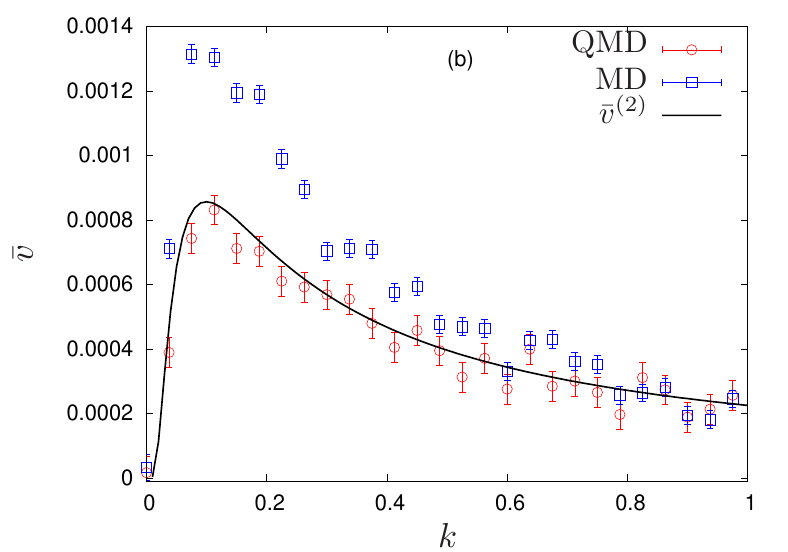}
\includegraphics[width=8cm]{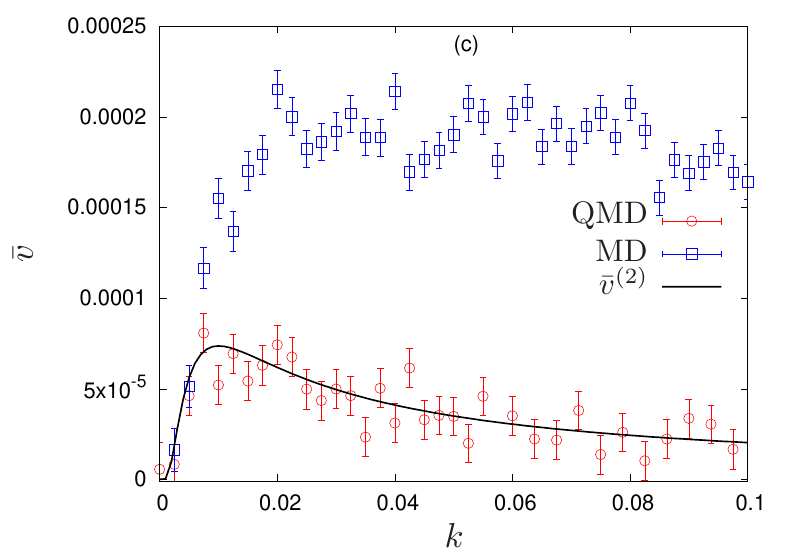}
\caption{Steady state center-of-mass velocity of the autonomous motor characterized by eq.~(\ref{H0:eq})--(\ref{U:quad}), as a function of the interaction strength $k$. The system parameters in reduced units read $\hbar=M=k_B =\eta_0=b=1$, $\vphi=\pi/2$, $T_1= \theta $, $T_2=2.5 \theta$. Panel (a): $\theta=1$, $V_0=0.5\theta$ and $V_0=0.75\theta$. Panel (b):  $\theta=0.1$, $V_0=0.75 \theta$. Panel (c) $\theta=0.01$, $V_0=\theta$ . Errorpoints: velocity obtained  through the QMD algorithm (circles, $10^4$ trajectories, $2^{22}$ time steps)  and the classical MD algorithm (squares, $10^5$ trajectories, $10^{6}$ time steps). Lines: second order velocity $\bar v^{(2)}$ as given by eq.~(\ref{v02}). }
\label{fig:QMD}
\end{figure}

The results are reported in fig.~\ref{fig:QMD}.
We notice that the agreement between the second order velocity $\bar v^{2}$ and the results obtained from the QMD is quite good, in the regime where the corrugation amplitude $V_0$ is smaller that the other energy scale (i.e. the thermal energy $k_B T$ and the energy scale associated with the harmonic interaction $\hbar \sqrt{k/m}$). For completeness we also plot the steady state velocity as obtained from the classical molecular dynamics (MD) algorithm.

\begin{figure}[h]
\center
\psfrag{v}[ct][ct][1.]{$\bar v_F$}
\psfrag{F}[ct][ct][1.]{$F$}
\psfrag{QMD}[lb][lb][.9]{QMD}
\psfrag{MD}[lb][lb][.9]{MD}
\psfrag{exact}[lc][lc][.9]{$\bar v^{(2)}_F$}
\includegraphics[width=8cm]{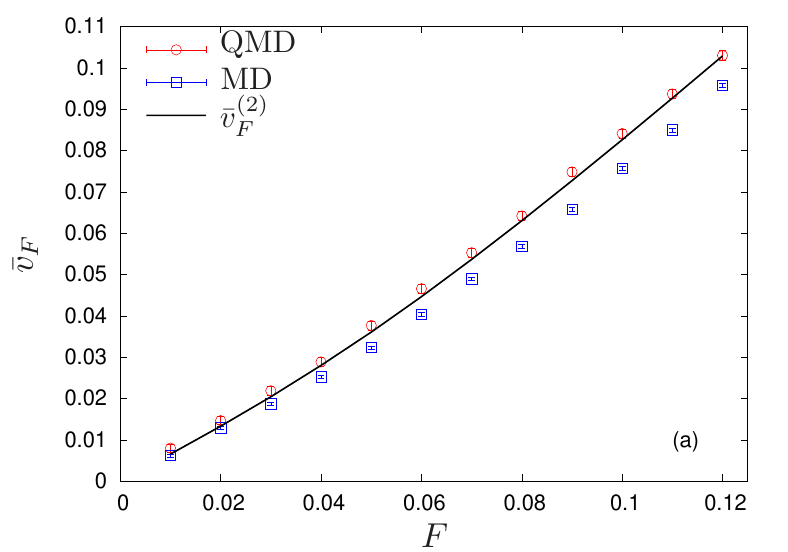}
\includegraphics[width=8cm]{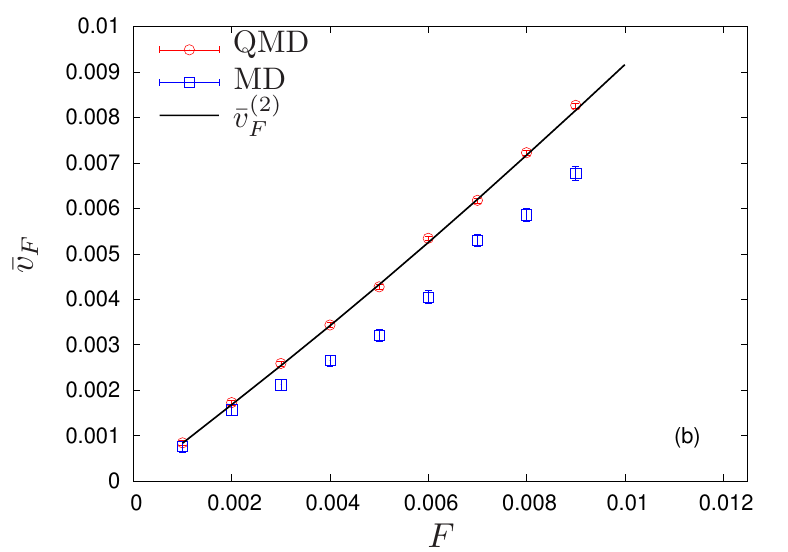}
\caption{Drift velocity $\bar v_F$ for a single particle in a potential $U(x)=-V_0 \cos(b x)-F x$ as a function of the force $F$. Full line theoretical prediction of ref.~\cite{Fisher85} for $\bar v_F^{(2)}$. Errorpoints: velocity as obtained through the QMD algorithm (circles, $10^3$ trajectories, $2^{17}$ time steps)  and the classical MD algorithm (squares, $10^3$ trajectories, $10^{6}$ time steps). The system parameters in reduced units read $\hbar=M=\eta_0=b=1$, $V_0=T=0.1$ (a), and $V_0=T=0.01$ (b). The external bath is Ohmic ($\Lambda \to \infty$)}
\label{fig:QMDF}
\end{figure}

As discussed in section~\ref{subsec:F}, when the interaction strength $k$ is set to zero, the model described by eqs. (\ref{lang1c})--(\ref{lang2c}) reduces to two independent particles moving under the effect of the external forces $F_i$.
This corresponds to the quantum Brownian particle in a
tilted sinusoidal potential $U(x)=-V_0 \cos(b x)-F x$, discussed in  \cite{Fisher85,Aslangul87}. In those references the second order velocity of the single particle has been calculated as a function of the applied force.
It is thus interesting to compare that exact results with the outcomes of the QMD algorithm in presence of an external constant force.
Such a comparison is shown in fig.~\ref{fig:QMDF} for two different choices of the parameter set: the agreement between the expected curve and the velocity predicted by the QMD algorithm is quite good.

\begin{figure}[h]
\center
\psfrag{v}[ct][ct][1.]{$\bar v_F$}
\psfrag{F}[ct][ct][1.]{$F$}
\psfrag{QMD}[lb][lb][.9]{QMD}
\psfrag{MD}[lb][lb][.9]{MD}
\psfrag{exact}[lc][lc][.9]{$\bar v^{(2)}_F$}
\includegraphics[width=8cm]{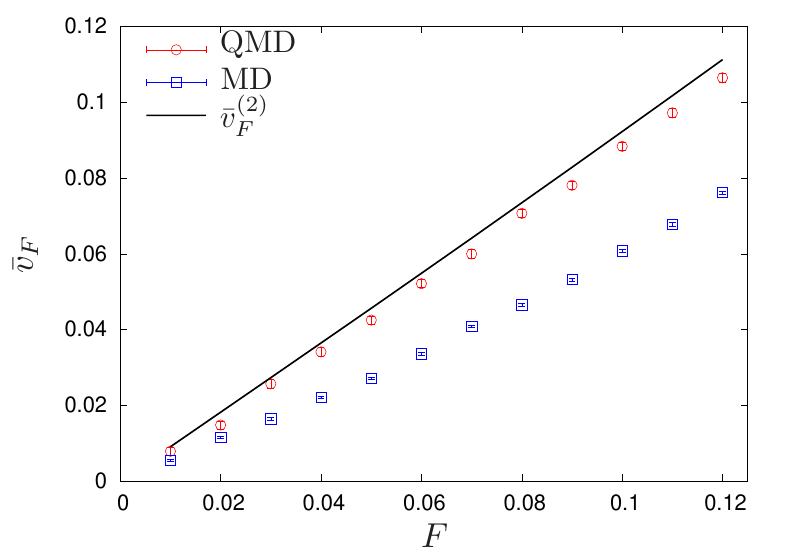}
\caption{Drift velocity $\bar v_F$ for a single particle in a potential $U(x)=-V_0 \cos(b x)-F x$ as a function of the force $F$, with $b=2$, all the other parameters are the same as in fig.~\ref{fig:QMDF} top. Full line theoretical prediction of ref.~\cite{Fisher85} for $\bar v_F^{(2)}$. Errorpoints: velocity as obtained through the QMD algorithm (circles, $10^3$ trajectories, $2^{17}$ time steps)  and the classical MD algorithm (squares, $10^3$ trajectories, $10^{6}$ time steps). We notice that decreasing the potential period worsens the agreement with the theoretical prediction, see discussion in the text.}
\label{fig:QMDF1}
\end{figure}

The agreement between the predicted second order velocity and the QMD algorithm worsens if one, e.g., reduces the potential period, see  fig.~\ref{fig:QMDF1}. 
This can be understood by noticing that the equilibrium QMD algorithm has already been reported to provide approximated results when tunnelling between neighbour wells become
predominant \cite{Dammak16}.

\section{Conclusions}
In this paper we have introduced and discussed the properties of a steady state quantum motor that can continuously convert heat flows into motion and thus work. As such the motor is different from the reciprocating motors performing thermodynamics cycles with "moving parts" which play the role of "pistons" in the classical picture, as discussed, e.g, in \cite{Alicki18}.

The present model relies only on the broken spatial symmetry of the underlying potentials  and on the temperature difference between the two baths. 
\alb{Had one considered the interaction potential $U(x_1-x_2)=-k \cos[b (x_1-x_2)]$ discussed in section~\ref{model:sec}, the system total potential (\ref{V:eq}) would be periodic, and so the expected velocity would be} the same if one considers periodic tracks (the two potentials $V_i(Q)$) with periodic boundary conditions, where the system only can move in the interval $[0, 2 \pi/b]$, or periodic tracks with open boundary conditions, where the system  can move in the range  $\left] -\infty, +\infty\right[$. \alb{With such a cosine potential our model would be similar to the quantum rotating gears discussed in, e.g., \cite{MacKinnon2002,Liu19}.  The quadratic potential used through this paper, eq.~(\ref{U:quad}), represents the large coupling limit (large $k$) of such a cosine interaction potential. Therefore in this limit, characterized by small fluctuations of the relative coordinate $y$, one should also expect that the steady state velocity is the same for the two different boundary conditions.}

Experimentally, there are quite a few examples in literature of systems  moving in periodic potentials and subject to temperature gradients.
In the context of classical stochastic thermodynamics, experiments where a single particle is constrained on periodic circular tracks have  been performed, e.g, in 
\cite{BlickleSpeck07} where a 3D toroidal laser trap was used to force a colloidal particle to perform circular trajectories along a periodic potential.
Brownian systems with 2 degrees of freedom, where the local temperature can be controlled, have been studied, for example, in \cite{Ciliberto2013,Ciliberto2013a,Berut16,Berut16a}.

The quantum regime considered here should potentially be realizable with two sideband laser-cooled atomic ions commonly confined by a harmonic potential, while individually interacting with each their periodic dipole-induced potential with variable spatial phase relation \cite{Linnet12,Enderlein2012,Bylinskii2015,Laupretre2019, Linnet2014}. The individual temperatures of the two ions can be controlled by addressing each of the ions with specific laser cooling beams. For single ions, temperatures of few $\mu$K have been achieved \cite{Poulsen2012}, and even tens of mK deep periodic dipole-induced potentials have been applied to smaller strings of ions \cite{Linnet12}. In this ion-scenario the common harmonic potential will eventual impede the ions’ motion, and instead lead to a stationary offset in the positions of the ions as compared to the thermalized case without the periodical potential. Moving the center of the common trapping potential with a constant velocity corresponding to the relevant one in fig.~\ref{figb} should, however, not lead to any displacement of the ions in the moving trap frame.


Our results on the QMD provide a solid evidence that such an algorithm can be successfully used to predict dynamical properties, such as particle currents, in quantum, out-of-equilibrium systems in contact with multiple reservoirs.
It would be interesting to test the algorithm on other systems whose dynamic properties are known exactly.

\alb{Finally, it would be interesting to compare the exact results discussed in this paper with those obtained through a master equation approach, corresponding to a weak-coupling  description of the interaction with the baths \cite{Breuer02}. Indeed, it has been found that the local or global master equations may give different dynamics in systems in contact with baths at different temperatures \cite{Hovhannisyan_2019}.}

\begin{acknowledgments}
This work was supported by the Danish Council for Independent Research and the Villum Foundation.
The numerical  results presented in this work were obtained at the Centre for Scientific Computing, Aarhus http://phys.au.dk/forskning/cscaa.
We thank F. Barra  for a critical reading of the manuscript, and helpful comments.
AI is grateful to Hans C. Fogedby for several useful discussions.
\end{acknowledgments}

\bibliography{bibliography}

\newpage

\appendix
\section{Two time correlation functions}
\label{app1}
In this appendix we calculate the two time correlations for the zeroth order operators $\average{\qo_i(t)\qo_j(t')}$.
From eqs.~(\ref{q1om})--(\ref{q2om}) in the main text we obtain 
\begin{widetext}
\begin{eqnarray}
\average{\qo_1(t)\qo_1(t')}=b^2\int \frac{\D \omega}{2 \pi} \E^{-\ii \omega(t-t')}\frac  {N(\omega)}{4}&&\left[\tilde F_1(\omega) \p{|\tilde G_x(\omega)|^2 +|\tilde G_y(\omega)|^2+ 2 \Re(G_x(\omega)G_y(-\omega))}\right.\nonumber\\
&& \left.+\tilde F_2(\omega) \p{|\tilde G_x(\omega)|^2 +|\tilde G_y(\omega)|^2- 2 \Re(G_x(\omega)G_y(-\omega))}\right],\\
\average{\qo_2(t)\qo_2(t')}=b^2\int \frac{\D \omega}{2 \pi} \E^{-\ii \omega(t-t')}\frac  {N(\omega)}{4}&&\left[\tilde F_1(\omega) \p{|\tilde G_x(\omega)|^2 +|\tilde G_y(\omega)|^2- 2 \Re(G_x(\omega)G_y(-\omega))}\right.\nonumber\\
&& \left.+\tilde F_2(\omega) \p{|\tilde G_x(\omega)|^2 +|\tilde G_y(\omega)|^2+ 2 \Re(G_x(\omega)G_y(-\omega))}\right],\\
\average{\qo_1(t)\qo_2(t')}=b^2\int \frac{\D \omega}{2 \pi} \E^{-\ii \omega(t-t')}\frac  {N(\omega)}{4}&&\left[\tilde F_1(\omega) \p{|\tilde G_x(\omega)|^2 -|\tilde G_y(\omega)|^2- 2 \ii \Im(G_x(\omega)G_y(-\omega))}\right.\nonumber\\
&& \left.+\tilde F_2(\omega) \p{|\tilde G_x(\omega)|^2 -|\tilde G_y(\omega)|^2+ 2 \ii \Im(G_x(\omega)G_y(-\omega))}\right],\\
\average{\qo_2(t)\qo_1(t')}=b^2\int \frac{\D \omega}{2 \pi} \E^{-\ii \omega(t-t')}\frac  {N(\omega)}{4}&&\left[\tilde F_1(\omega) \p{|\tilde G_x(\omega)|^2 -|\tilde G_y(\omega)|^2+ 2 \ii \Im(G_x(\omega)G_y(-\omega))}\right.\nonumber\\
&& \left.+\tilde F_2(\omega) \p{|\tilde G_x(\omega)|^2 -|\tilde G_y(\omega)|^2- 2 \ii \Im(G_x(\omega)G_y(-\omega))}\right].
\end{eqnarray} 
We notice that the same time correlation functions $\average{\qo_i(t)\qo_j(t)}$ are independent of the time.

One can thus calculate the two-time two-particle correlation function (\ref{C12}) in the main text, and the single particle correlation functions $c_{ii}(\tau)$ appearing in eqs.~(\ref{q12F})--(\ref{IF:def}) 
\begin{eqnarray}
c_{11 (22)}(t-t')&=&\average{(\qo_{1 (2)}(t)-\qo_{1 (2)}(t'))^2}=\nonumber\\
&=&\frac{b^2} 2 \int \frac{\D \omega}{2 \pi}N(\omega)(1-\cos \omega(t-t'))\left\{\tilde F_1(\omega)
 \pq{|\tilde G_x(\omega)|^2 +|\tilde G_y(\omega)|^2 \pm  2 \Re(G_x(\omega)G_y(-\omega))} \right. \nonumber \\
&&\left. \qquad \qquad \qquad  \qquad \qquad  \qquad  \qquad \quad  \tilde F_2(\omega)
 \pq{|\tilde G_x(\omega)|^2 +|\tilde G_y(\omega)|^2 \mp 2 \Re(G_x(\omega)G_y(-\omega))}\right\} .
\label{C11}
\end{eqnarray} 
\end{widetext}

\section{Taylor expansion of the sinusoidal functions}
\label{app1a}
In this appendix we derive the expansion of the trigonometric functions on the rhs of eq.~(\ref{aveX}) up to the first order in $V_0$.
By writing the trigonometric functions in their exponential forms, and properly differentiating the exponential operators (see appendix ~\ref{app2})  one obtains for the sine
\begin{eqnarray}
&&\pg{\sin (\qo(t)+V_0 \qu(t))}^{(1)}=\nonumber\\
&&\p{\qu-\frac{1}{3!} [\qo,[\qo,\qu]]+\dots}\cos \qo   \nonumber\\
&&+\p{-\frac{1}{2!}[\qo,\qu]+\frac{1}{4!}[\qo,[\qo,[\qo,\qu]]]+\dots }\sin \qo\nonumber\\
\label{sinexp:0}
\end{eqnarray}
By setting $\hqo_2=\qo_2+\pi/2$, one obtains the expansion for the cosine function in eq.~(\ref{aveX}).

We have now to evaluate the series of nested commutators in (\ref{sinexp:0}). One can do this by noticing that the operators $\qu_1$ and $\qu_2$ are linear functionals of  $\sin \qo_1(t)$ and $\cos \bar q^{0}_2(t)$, see eqs.~(\ref{q11})-(\ref{Y1}) and appendix (\ref{app2}).
Thus one obtains  
\begin{widetext}
\begin{eqnarray}
&&\pg{\sin (\qo(t)+V_0 \qu(t))}^{(1)}=\nonumber\\
&&=-\frac {b^2}2 \int_{t_0}^t \D\tau \, (G_x(t-\tau)+G_y(t-\tau))\pq{\Sigma_{11} (\tau-t)\sin \qo_1(\tau) \cos \qo_1(t) + \Sigma'_{11} (\tau-t)\cos q^{(0)}_1(\tau) \sin \qo_1(t)} +\nonumber\\
&& \qquad\qquad\qquad   (G_x(t-\tau)-G_y(t-\tau))\pq{\Sigma_{21} (\tau-t)\sin  \hqo_2(\tau) \cos \qo_1(t) + \Sigma'_{21} (\tau-t)\cos \hqo_2(\tau) \sin \qo_1(t)}\label{sinexp}\\ 
&&\pg{\cos (\qo_2+V_0\qu_2)}^{(1)}=\pg{\sin (\hqo_2+V_0\qu_2)}^{(1)}=\nonumber\\
&&=-\frac {b^2} 2 \int_{t_0}^t \D\tau \, (G_x(t-\tau)-G_y(t-\tau))\pq{\Sigma_{12} (\tau-t)\sin \qo_1(\tau) \cos \hqo_2 (t) + \Sigma'_{12} (\tau-t)\cos q^{(0)}_1(\tau) \sin \hqo_2(t)} +\nonumber\\
&& \qquad\qquad\qquad   (G_x(t-\tau)+G_y(t-\tau))\pq{\Sigma_{22} (\tau-t)\sin  \hqo_2(\tau) \cos \hqo_2(t) + \Sigma'_{22} (\tau-t)\cos \hqo_2(\tau) \sin \hqo_2(t)}\label{cosexp},
\end{eqnarray}
\end{widetext}
where 
\begin{eqnarray}
\Sigma_{ij}&=&1 +\frac{\p{2 a_{ij}}^2}{3 !}+\frac{\p{2 a_{ij}}^4}{5 !}+\dots=\frac{\cos a_{ij}\sin a_{ij}}{a_{ij}},\nonumber\\
\Sigma'_{ij}&=&\frac{-2 i a_{ij}}{2 !}+\frac{\p{-2i  a_{ij}}^3}{4 !}+\dots=-i \frac{\sin^2  a_{ij}}{a_{ij}},\nonumber
\end{eqnarray} 
and where the function $a_{ij}(t-t')$ has been defined in eq.~(\ref{def:aij}), and is proportional to the commutator of the position operators at different times, which are {\it c}-numbers as discussed above. Therefore the functions $\Sigma_{ij}(t-t')$ and $\Sigma'_{ij}(t-t')$ are also complex-valued functions of the time lapse $t-t'$.
We furthermore notice that the following equality holds
\begin{equation}
\Sigma_{ij}\pm \Sigma'_{ij}=\frac{\E^{\mp i a_{ij}}}{a_{ij}} \sin a_{ij}.
\label{DS}
\end{equation} 
Our goal is to calculate the average on the right hand side of eq.~(\ref{aveX}), and we recall that the average has to be taken over the bath variables, represented by the noise operators $\xi_i$: such operators appear in the zeroth-order solutions for the position operators $\qo_i(t)$, see eqs.~(\ref{X0})-(\ref{Y0}).
Thus we have to calculate the average over the bath variables of the products of trigonometric functions appearing on the rhs of eqs.~(\ref{sinexp})--(\ref{cosexp}). 
It is convenient to express those sine and cosine functions in their complex exponential form.
Furthermore we notice once more that the commutators $\pq{q_i(t),q_j(t')}$ are  {\it c}-numbers, so we can use the Glauber formula
\begin{equation}
\E^{\ii \alpha q_i(t)} \E^{\ii \alpha' q'_i(t')}=\E^{\ii (\alpha q_i(t)+\alpha' q'_i(t'))} \E^{-\frac 1 2 \alpha\alpha' \pq{q_i(t),q_j(t')}},
\end{equation}  
with $\alpha,\, \alpha'=\pm 1$.
The zero-th order particle positions $\qo_i(t)$ are linear combinations of the bath variables, which are Gaussian variables with zero average.
We can therefore use the following 
equality
\begin{equation}
\langle \E^{\ii \p{\qo_i(t)\pm\qo_j(t')}}\rangle=\E^{-\frac 1 2  \average{(\qo_i(t)\pm\qo_j(t'))^2}}
\end{equation} 

Thus a straightforward calculation  gives
\begin{eqnarray}
&&\average{\sin \hqo_2(\tau'') \cos \qo_1(t)}=\nonumber\\
&&\, =\frac 1 2 \left( \E^{-\frac 1 2 \average{\p{ \qo_2(\tau'') + \qo_1(t)}^2} - \frac 1 2 \pq{\qo_2(\tau'') , \qo_1(t)}} \right.\nonumber \\
&&\qquad \left. + \E^{-\frac 1 2 \average{\p{ \qo_2(\tau'') - \qo_1(t)}^2} + \frac 1 2 \pq{\qo_2(\tau'') , \qo_1(t)}} \right)\label{s2c1}
\end{eqnarray}
\begin{eqnarray}
&&\average{\cos \hqo_2(\tau'') \sin \qo_1(t)}=\nonumber \\
&&\,=\frac 1 2 \left( \E^{-\frac 1 2 \average{\p{ \qo_2(\tau'') + \qo_1(t)}^2} - \frac 1 2 \pq{\qo_2(\tau'') , \qo_1(t)}}\right.\nonumber\\
&&\qquad \left.- \E^{-\frac 1 2 \average{\p{ \qo_2(\tau'') - \qo_1(t)}^2} + \frac 1 2 \pq{\qo_2(\tau'') , \qo_1(t)}} \right)\label{c2s1}
\end{eqnarray}
\begin{eqnarray}
&&\average{\sin  \qo_1(\tau'') \cos \hqo_2(t)}=\nonumber\\
&&\, = \frac 1 2 \left( \E^{-\frac 1 2 \average{\p{ \qo_1(\tau'') + \qo_2(t)}^2} - \frac 1 2 \pq{\qo_1(\tau'') , \qo_2(t)}}\right.\nonumber\\
&&\qquad \left. -\E^{-\frac 1 2 \average{\p{ \qo_1(\tau'') - \qo_2(t)}^2} + \frac 1 2 \pq{\qo_1(\tau'') , \qo_2(t)}} \right)\label{s1c2} \\
&&\average{\cos  \qo_1(\tau'') \sin \hqo_2(t)}=\nonumber\\
&& \, =\frac 1 2 \left( \E^{-\frac 1 2 \average{\p{ \qo_1(\tau'') + \qo_2(t)}^2} - \frac 1 2 \pq{\qo_1(\tau'') , \qo_2(t)}}\right.\nonumber \\
&&\qquad \left.+\E^{-\frac 1 2 \average{\p{ \qo_1(\tau'') - \qo_2(t)}^2} + \frac 1 2 \pq{\qo_1(\tau'') , \qo_2(t)}}\right)\label{c1s2}.
\end{eqnarray}
We notice that the "even" terms $\langle \sin \qo_i(\tau) \cos \qo_i(t)\rangle$ in eqs.~(\ref{sinexp})--(\ref{cosexp}) vanish, as can be also checked by a direct calculation.

The variables $\qo_{1,2}=(\Xo\pm\Yo)/2$ depend on the variable $\Xo$ which describes a free Brownian motion.
As such the terms $\langle (\qo_i(\tau)+\qo_j(t))^2\rangle$ are divergent. Indeed they contain the term  $\langle (\Xo(\tau)+\Xo(t))^2\rangle$ and by using eqs.~(\ref{X0}), (\ref{q1om})--(\ref{q2om}) one finds 
\begin{eqnarray}
&&\average{(\Xo(t)\pm\Xo(t'))^2}=\nonumber\\
&& =b^2 \int \frac{\D \omega}{ \pi} N(\omega)|G_x(\omega)|^2  (\tilde F_1(\omega)+ \tilde F_2(\omega))(1\pm \cos \omega(t-t'))\nonumber\\
&&=b^2 \int \frac{\D \omega}{\pi} N(\omega)  \frac{(\tilde F_1(\omega)+ \tilde F_2(\omega))(1\pm \cos \omega(t-t'))}{\omega^2( m^2 \omega^2-2 m \omega \Im \tilde \eta(\omega)+ |\tilde \eta(\omega)|^2)}\nonumber,
\end{eqnarray} 
and the integrand with the plus sign on the right hand side diverges as $\sim1/\omega$ for $\omega \to 0$, given that $N(\omega=0)\neq0 $ as discussed in section \ref{model:sec}. Thus the terms $\exp\pq{ -\langle (\qo_i(\tau)+\qo_j(t))^2\rangle/2} $ in eqs. (\ref{s2c1})-(\ref{c1s2}) vanish.

By using the results in eqs.~(\ref{sinexp})--(\ref{cosexp}) and (\ref{s2c1})--(\ref{c1s2}), the right hand side of equation (\ref{aveX}) can thus be written as
\begin{eqnarray}
&&\int^t_{-\infty} \eta(t-t')\langle \dot x^{(2)}(t')\rangle \D t'=\nonumber\\
&&=\frac {b^4} 4  \int_{-\infty}^t \D\tau \, (G_x(t-\tau)-G_y(t-\tau)) \nonumber \\
&&\qquad \times \left[\frac{\sin a_{21}(\tau-t)}{ a_{21}(\tau-t)}\E^{-\frac 1 2 \average{( \qo_2(\tau) - \qo_1(t))^2}} \right.\nonumber \\
&& \qquad \quad\quad \left. -\frac{\sin a_{12}(\tau-t)}{ a_{12}(\tau-t)} \E^{-\frac 1 2 \average{( \qo_1(\tau) - \qo_2(t))^2}}\right],\nonumber\\
&&=\frac {b^4} 4  \int_{-\infty}^t \D\tau \, (G_x(t-\tau)-G_y(t-\tau)) \nonumber \\
&&\qquad \qquad \times \frac{\sin a_{12}(\tau-t)}{ a_{12}(\tau-t)}\p{\E^{c_{21}(\tau-t)}-\E^{c_{12}(\tau-t)}}\nonumber\\
&&=b^2 \mathcal I
\label{aveX1}
\end{eqnarray} 
where
\begin{eqnarray} 
&&\mathcal I=\frac 1 4  \int_{0}^{+\infty} \D\tau \, (G_x(\tau)-G_y(\tau)) \nonumber \\
&&\qquad \qquad \frac{\sin a_{12}(\tau)}{ A_{12}(\tau)}\p{\E^{-\frac 1 2 c_{12}(\tau)}-\E^{-\frac 1 2 c_{21}(\tau)}}
\label{defIapp}
\end{eqnarray} 
and where we have used eq.~(\ref{DS}).

By noticing that the last term in eq.~(\ref{aveX1}) is time-independent, and by recalling the definition of the memory function (\ref{aeta1}), one can invert equation~(\ref{aveX1}) and obtain the expression for the long time steady state velocity eq.~(\ref{avex2}) in the main text.

\section{Taylor expansion of eq.~(\ref{sinexp:0})}
\label{app2}
In order to prove eq.~(\ref{sinexp:0}) in the previous appendix we need to calculate the first order contribution to the expression 
\begin{equation}
\sin (\qo(t)+V_0 \qu(t)).
\label{sinapp}
\end{equation} 
The calculation is more easily performed when the sine is expressed in terms of complex exponentials $\exp(\pm \ii(\qo(t)+V_0 \qu(t))$.  Where in general the operators $\qo(t)$ and $\qu(t)$ do not commute.

Let $X$ and $Y$ be two non commuting operators, and let us introduce the operator $A(\epsilon)=X+\epsilon Y$, for which the operator identity holds \cite{Snider64}
\begin{eqnarray}
\frac{\mathrm d}{{\mathrm d} \epsilon}\E^{A(\epsilon)}=\int_0^1\D x\, \E^{x A(\epsilon)} \frac{\mathrm dA(\epsilon)}{{\mathrm d} \epsilon}\E^{-x A(\epsilon)}  \E^{ A(\epsilon)}.
\end{eqnarray} 
Thus one has \cite{Feynman51,Snider64} 
\begin{eqnarray}
&&\left.\frac{\mathrm d}{{\mathrm d} \epsilon}\E^{A(\epsilon)}\right|_{\epsilon=0}=\nonumber \\
&=&(Y+\frac{1}{2!}[X,Y]+\frac{1}{3!}[X,[X,Y]]+\dots) \E^{X}.
\label{dAdx0}
\end{eqnarray} 
By taking $A(\epsilon)=\pm \ii (\qo(t)+\epsilon \qu(t))$ and by applying the above result (\ref{dAdx0}), one obtains the first order term of eq.~(\ref{sinapp}), and thus eq.~(\ref{sinexp:0}) in the main text.

\section{Proof of the equality $\average{\ddot  x^{(2)}}=0$}
\label{app1ax}
In order to prove that in the steady state the average center-of-mass velocity is constant we proceed as follows.
By summing eqs.~(\ref{lang12}) and (\ref{lang22}), we obtain
\begin{eqnarray}
M\ddot x^{(2)}&=&-\int_{-\infty}^t \eta(t-t') \dot x^{(2)}(t') \D t'-b^2 h(t)\\
h(t)&=&  \pg{\sin (\qo_1+ V_0\qu_1)}^{(1)}+ \pg{\cos (\qo_2+ V_0\qu_2)}^{(1)}\nonumber \\
&& 
\end{eqnarray}
whose solution reads
\begin{equation}
x^{(2)}(t)=
-b^2 \int_{-\infty}^{t} G_x(t-t') h(t') \D t',
 \end{equation} 
and where the Fourier transform of $ G_x(t)$ is given by eq.~(\ref{Gx:def}).
The second derivative reads 
\begin{equation}
\ddot  x^{(2)}= -b^2 G_x(0)\partial_t h(t)+ b^2 \int_{-\infty}^t \partial_{t'} G(t-t') \partial_{t'}h(t')\D t'.
\label{eq:ddotx2}
\end{equation} 
By averaging over the quantum noise, we thus obtain
\begin{equation}
\average{ \ddot  x^{(2)}}= -b^2 G_x(0)\average{\partial_t h(t)}+ b^2 \int_{-\infty}^t \partial_{t'} G(t-t')\average{ \partial_{t'}h(t')}\D t'.
\end{equation} 
The average $-b^2 \average{h(t)}$ is given by the rhs of eq.~(\ref{aveX1}), see also eq.~(\ref{aveX}). Therefore we have $-b^2 \average{h(t)}=b^2 \mathcal I$,
where $\mathcal I$ is given by eq.~(\ref{defIapp}) and is time independent. Thus, as the systems is in the steady state
$\average{ \partial_{t}h(t)}=\partial_{t}\average{ h(t)}=0$, and the rhs of eq.~(\ref{eq:ddotx2}) vanishes, which is the desired  result.

\end{document}